\documentclass[pra,twocolumn,showpacs,superscriptaddress,floatfix, nofootinbib]{revtex4-2}

\usepackage{siunitx}

\usepackage{amssymb}
\usepackage{amsmath}
\usepackage{hyperref}
\usepackage{graphicx}
\usepackage{array}

\usepackage{multirow}

\usepackage{tabularx}
\usepackage{ragged2e}
\usepackage{comment}

\usepackage{physics}
\usepackage{braket}

\usepackage[autostyle,english=american]{csquotes}
\MakeOuterQuote{"}

\usepackage[table,xcdraw]{xcolor}
\usepackage[normalem]{ulem}

\usepackage{makecell}
\usepackage{booktabs}
\usepackage{subcaption}
\begin{document}

\title{Proposal for a multi-ion optical clock with hundreds of ions}

\author{Pavel A. Kamenskikh}
\affiliation{P.N. Lebedev Physical Institute of the Russian Academy of Sciences, Moscow 119991, Russia}
\author{\textcolor{black}{Vasiliy N. Smirnov} }
\affiliation{P.N. Lebedev Physical Institute of the Russian Academy of Sciences, Moscow 119991, Russia}

\author{Ilya S. Gerasin}
\affiliation{P.N. Lebedev Physical Institute of the Russian Academy of Sciences, Moscow 119991, Russia}

\author{Ilya A. Semerikov}
\affiliation{P.N. Lebedev Physical Institute of the Russian Academy of Sciences, Moscow 119991, Russia}
\affiliation{
Key Laboratory of Quantum State Construction and Manipulation (Ministry of Education), School of Physics, Renmin University of China, Beijing 100872, China}
\affiliation{Russian Quantum Center, Skolkovo, Moscow 121205, Russia}

\author{Wei Zhang}
\affiliation{
Key Laboratory of Quantum State Construction and Manipulation (Ministry of Education), School of Physics, Renmin University of China, Beijing 100872, China}
\affiliation{Russian Quantum Center, Skolkovo, Moscow 121205, Russia}
\affiliation{
Beijing Academy of Quantum Information Sciences, Beijing 100193, China}

\begin{abstract}
Scaling trapped-ion optical clocks to hundreds of ions remains a major challenge due to inhomogeneous frequency shifts in large ion crystals. Here, we propose a circular surface-trap architecture for a multi-ion optical clock with hundreds of ions. The rotational symmetry of the ion crystal suppresses the main sources of inhomogeneity, quadrupole shift.We investigate the feasibility of a clock based on a ring crystal of 400~$^{40}\mathrm{Ca}^{+}$ ions and propose a suitable experimental setup configuration.  We have shown that circular configuration of ions leads to suppression by two orders of magnitude of the quadrupole shift comparing to the linear trap. Combined with stray-field compensation and dynamical decoupling techniques, this architecture provides a pathway toward multi-ion optical clocks with fractional frequency uncertainty at the $10^{-18}$ level and an expected 20-fold improvement in frequency stability compared to single ion clocks.
\end{abstract}

\maketitle 

\section{Introduction}\label{sec1}

Trapped ions represent one of the most accurate and systematically controlled platforms for optical frequency metrology. Recent advances in laser stabilization, environmental control, and systematic-shift evaluation have enabled fractional systematic uncertainties below the $10^{-18}$ level in several ion species, including $^{27}\mathrm{Al}^{+}$, $^{40}\mathrm{Ca}^{+}$, $^{88}\mathrm{Sr}^{+}$ and $^{171}\mathrm{Yb}^{+}$  \cite{brewer201927,sanner2019optical,marshall2025HighStability,zhang2026LiquidNitrogenCooled, filzinger2026multiion}. Such performance has established trapped-ion clocks as key tools for relativistic geodesy, searches for ultralight dark matter, tests of Lorentz symmetry, and investigations of temporal variations of fundamental constants~\cite{safronova2018search, mehlstaubler2018Atomic, safronova2019search, sherrill2023analysis, chu2025Exploring}. 
  
In spite of the outstanding accuracy, most state-of-the-art ion clocks operate with a single interrogated ion. Consequently, their short-term stability is fundamentally limited by quantum projection noise. In contrast, neutral-atom optical lattice clocks routinely interrogate ensembles containing ($10^3-10^5$) atoms and therefore achieve substantially improved measurement statistics~\cite{bloom2014optical, ludlow2015optical, mcgrew2018atomic, young2020Halfminutescale, aeppli2024Clock, aeppli2025atomic}. Even for the most stable single-ion clocks, averaging times approaching one day are required to achieve statistical uncertainties at the $10^{-18}$ level~\cite{marshall2025HighStability, zhang2026LiquidNitrogenCooled}. Increasing the number of simultaneously interrogated ions therefore represents a natural way toward improving clock stability. Large ion ensembles are also attractive for the implementation of cascaded clock architectures~\cite{borregaard2013efficient} and entanglement-enhanced spectroscopy protocols capable of exceeding the standard quantum limit~\cite{leibfried2004HeisenbergLimited, kielinski2024GHZ, korolkov2025Multiqubit}. 

However, scaling trapped-ion clocks beyond a few particles remains a major challenge due to inhomogeneous frequency shifts, which cause spectral line broadening and therefore reduce the precision of multi-ion clocks~\cite{arnold2015Prospects, aharon2019Robust}. The dominant inhomogeneous shifts originate from electric and magnetic fields. An electric field gradient coupled to the quadrupole moment of the clock state induces a quadrupole shift~\cite{itano2000Externalfield}. Another dominant contribution arises from magnetic-field fluctuations, which is especially important for ions without hyperfine splitting, such as~$^{40}\mathrm{Ca}^{+}$ and~$^{88}\mathrm{Sr}^{+}$.
 
 Several methods have been developed to suppress magnetic-field effects, including magnetic shielding and active stabilization~\cite{merkel2019Magnetic, ruster2016longlived}. Elimination of quadrupole shifts requires additional treatment. For a single ion, the quadrupole shift can be suppressed by optimizing the magnetic-field orientation~\cite{tan2019suppressing, keller2019Controlling, filzinger2026multiion}, averaging over different magnetic-field directions~\cite{lange2020Coherent}, or averaging over multiple Zeeman transitions~\cite{dube2005Electric, chwalla2009Absolute, dube2013Evaluation, oskay2005Measurement}. However, spatially varying fields lead to ion-dependent shifts, making the averaging schemes ineffective. Suppression of the quadrupole shift by controlling the magnetic-field angle was recently demonstrated in a multi-ion clock with eight~$^{88}\mathrm{Sr}^{+}$ ions~\cite{filzinger2026multiion}, although further scaling of this approach remains challenging.
Other strategies to overcome these effects in multi-ion clocks include selecting ion species with a negligible quadrupole moment such as~$^{115}\mathrm{In}^{+}$ or~$^{27}\mathrm{Al}^{+}$~\cite{herschbach2012Linear, keller2019Controlling, hausser2025115} and the use of continuous dynamical decoupling techniques, which average over multiple magnetic sublevels and simultaneously suppress inhomogeneous quadrupole and Zeeman shifts~\cite{aharon2019Robust, shaniv2019Quadrupole, kaewuam2020Hyperfine, pelzer2024Multiion, akerman2025Operating}.

Despite the aforementioned challenges, recent experimental progress has demonstrated the feasibility of precision spectroscopy in multi-ion systems.  Filzinger \emph{et al.} demonstrated an optical clock based on eight simultaneously interrogated $^{88}\mathrm{Sr}^{+}$ ions with a total systematic uncertainty of ($5\times10^{-19}$) \cite{filzinger2026multiion}. Earlier multi-ion clock studies with $^{88}\mathrm{Sr}^+$ ions~\cite{akerman2025Operating, steinel2023Evaluation} and~$^{115}\mathrm{In}^+$ ions~\cite{hausser2025115} should be emphasized. All of these multi-ion clock demonstrators utilize a linear ion chain configuration which is exceptionally difficult to scale toward the hundreds-of-ions regime.

In this work, we explore a different architecture: a microfabricated ring surface trap capable of symmetrically confining a large Coulomb crystal containing hundreds of ions with nearly uniform spacing. Such traps were originally developed for quantum computing architectures~\cite{tabakov2015Assembling} and are used in some state-of-the-art devices~\cite{ransford2025Helios}. An alternative approach to realizing ring-shaped crystals is the use of multipole traps~\cite{okada2007Crystallization, clark2013Ideal}, which have been proposed as an architecture for multi-ion optical clocks~\cite{champenois2010Ion}. Here, we instead propose a segmented ring surface trap architecture for multi-ion clocks and analyze the dominant inhomogeneous shift arising in this scheme.

The key feature of the ring geometry — the rotational symmetry of the ion crystal — can be exploited for multi-ion optical clock operation, as it naturally suppresses the quadrupole shift and reduces the effect of micromotion. The trade-off of this geometry is that stray electric fields can distort the ion crystal, introducing additional inhomogeneities that degrade clock accuracy~\cite{tabakov2015Assembling}. We address this challenge using established stray-field compensation techniques~\cite{tabakov2015Assembling}.
We further propose an experimental setup configuration suitable for implementation of a multi-ion clock with hundreds of ions, including a readout scheme capable of resolving a 1-mm-scale ion crystal and a flat-top laser-beam profiles to ensure homogeneous interrogation across the entire crystal.

\section{Experimental Architecture}
\begin{figure}[t]
\centering
\includegraphics[width=1\columnwidth]{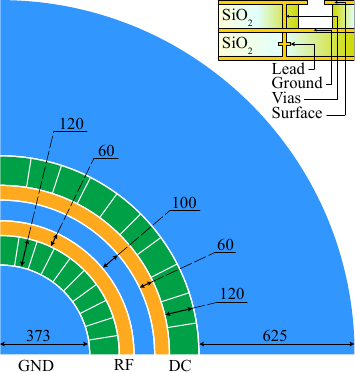}
\caption{\protect\justifying Schematic of the ring surface ion trap with dimensions of the ground (GND), radio-frequency (RF) and direct current (DC) electrodes indicated in micrometers. The gap size between all electrodes is 7 $\mu$m. The simplified layer structure is shown in the upper right corner.}
\label{ringtrap}
\end{figure}
In this section, we propose the architecture of the surface trap, readout and laser schemes, and discuss stray field compensation techniques for the multi-ion clock. The main challenges occur due to the size of the ion crystal and stray fields that break the circular symmetry of the ion chain.

\subsection{Ring surface trap architecture}
\begin{figure}[t]
\centering
\includegraphics[width=1\columnwidth]{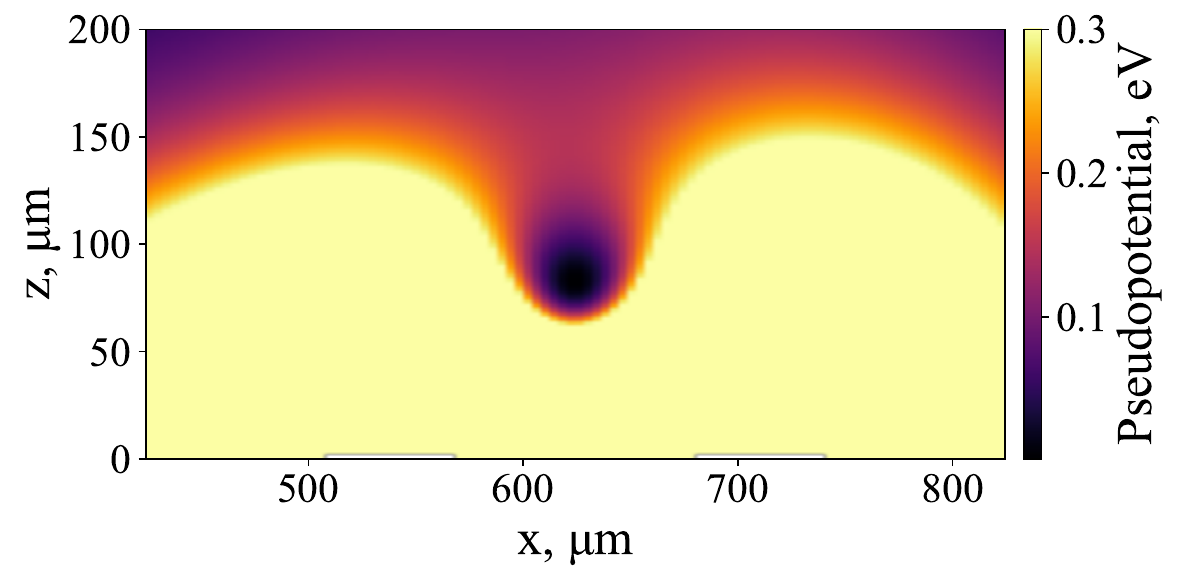}
\vspace{-18pt}
\caption{\protect\justifying Contour plot of the 0.15 eV deep pseudopotential of the ring trap achieved with 200 V amplitude at RF electrodes. }
\label{pseudo}
\end{figure}

We are inspired by the trap geometry and structure presented in~\cite{tabakov2015Assembling}. However, we consider different number $N \in \{25, 65, 100, 170\}$ pairs of compensating DC electrodes (one inner, one outer electrode per pair). All inner compensating DC electrodes have the same width, while the outer ones have the same geometry and the width adjusted accordingly. The dimensions of the trap are presented in Fig.~\ref{ringtrap}. The trap has an inner circular GND electrode with $r = 373 \text{ }\mu m$. The $N$ DC electrodes are arranged in a circle, all the gaps between electrodes are equal to $ g = 7 \text{ } \mu m$. The radial width of each DC electrode is $w_{dc} = 120 \text{ }\mu m$. Circular RF electrodes have a width of $w_{rf} = 60 \text{ }\mu m$. Intermediate circular GND electrode has a width of $w_{gnd} = 100 \text{ }\mu m$. Finally, there is an outer circular GND electrode with $w_{out} = 625 \text{ }\mu m$. The calculated pseudopotential of the trap with 0.1 eV depth is presented in Fig.~\ref{pseudo}. The calculations are made using Python-package electrode~\cite{electrode}. The RF-voltage amplitude is 200 V, and the driving frequency is 24.801 MHz, which is chosen to compensate for micromotion~\cite{berkeland1998Minimization, arnold2015Prospects}. The radial secular frequencies are $\{2.48,\, 2.51\}$ MHz. The axes of the trap are tilted by an angle of $\approx 13^\circ$ relative to the horizontal plane. This happens due to different areas of the inner and outer RF electrodes. The trap has upper metal layer with electrodes, and three other metal layers for routing and ground with a thick dielectric between them~\cite{tabakov2015Assembling}.

\vspace{-0.5 cm}
\subsection{Stray field compensation}
\label{sec:compensation}
The crucial requirement for a ring-trap multi-ion clock is the circular symmetry of the chain. However, as was demonstrated in Ref.~\cite{tabakov2015Assembling}, the circular symmetry of the ion crystal breaks due to the stray fields. Such undesired electric fields may occur from different sources, including the laser-induced charging of the trap surface, contaminants,  fabrication imperfections, and other uncontrolled charges in the vacuum chamber~\cite{tabakov2015Assembling, clark2014Characterization}. With the tangential stray field compensation method proposed in~\cite{tabakov2015Assembling}, the circular symmetry of the ion crystal in the ring trap can be recovered. 

In linear ion chains, stray electric fields lead to ion-dependent quadrupole shifts. These shifts are commonly compensated by applying DC voltages to segmented electrodes, with the required corrections determined from micromotion measurements~\cite{keller2019Controlling}. The central idea of the present work is that an equidistant circular ion crystal can significantly suppress the spatial variation of the quadrupole shift. Achieving this, however, requires compensating the stray electric field at the position of every ion. Assuming a chain of 400 ions, this corresponds to enforcing field compensation at 400 equally spaced locations. A straightforward implementation would require approximately three independently controlled electrodes per ion (to control the three field components), resulting in roughly 1200 electrodes. Such a large number of electrodes considerably complicates the chip layout and electrical routing. Moreover, this number cannot be placed with the gap size equal to 7 $\mu$m, which is a common technological limitation. For our geometry, the maximum number of electrode pairs is equal to $N= 170$. We therefore investigated the minimum number of control electrodes required to realize an equidistant circular ion chain.

For a fixed number of electrode pairs $N$, we define 400 uniformly distributed target points along the RF null, corresponding to the desired ion positions. At each point $i$, the normal, tangential, and z electric-field contribution from each control electrode $n$ $\{E^n_{i,n}, E_{i,n}^t, E_{i,n}^z\}$ is calculated. As a representative example of experimental imperfections, we adopt the stray electric field distribution reported in Ref.~\cite{tabakov2015Assembling}. This assumption is justified because the trap geometry considered here closely resembles the one investigated there. Because the tangential component of the stray electric field, $E_i^{t}$, is reported at only 45 positions, we linearly interpolate the published data to obtain the electric-field distribution at all 400 target points. The normal and vertical components, ${E_i^n}$ and ${E_i^z}$, are therefore assumed to be zero throughout this study. The compensation voltages $\{V_n\}$ are then obtained by solving the resulting underdetermined system of linear equations:

\begin{equation}
    \begin{pmatrix}
    E^n_{1, 1} & ... & E^n_{1, 2N}\\
    ...        & ... & ... \\
    E^n_{400, 1} & ... & E^n_{400, 2N}\\
    E^t_{1, 1} & ... & E^t_{1, 2N}\\
    ...        & ... & ... \\
    E^t_{400, 1} & ... & E^t_{400, 2N}\\
    E^z_{1, 1} & ... & E^z_{1, 2N}\\
    ...        & ... & ... \\
    E^z_{400, 1} & ... & E^z_{400, 2N}\\
    \end{pmatrix} 
    \begin{pmatrix}
    V_1 \\ ... \\ V_{2N}
    \end{pmatrix} 
    = -
    \begin{pmatrix}
    0 \\ ... \\ 0 \\ {E^t_1} \\ .. \\ {E^t_{400}} \\ 0 \\ .. \\ 0
    \end{pmatrix} 
\end{equation}

The first matrix we denote as $\boldsymbol{E_b}$, the vector of voltages is $\vec{V}$, the vector of stray fields is denoted as $\vec{E_{s}}$. We use the least-squares solution to the underdetermined linear matrix equation with regularization to limit maximal voltages (normalization coefficient $\lambda$ is equal to $10^{-4}$):
\begin{equation}
    \min_{V_i} ||\boldsymbol{E_{b}} \vec{V} + \vec{E_{s}} || + \lambda ||\vec{V}||.    
\end{equation}
Restricting the analysis to a single field component does not affect the generality of the approach, since the compensation problem remains an inhomogeneous linear system. Including the remaining field components would increase the number of non-zero elements on the right-hand side but would not alter the mathematical formulation of the problem.

\begin{figure*}[t]
\centering
\includegraphics[width=2\columnwidth]{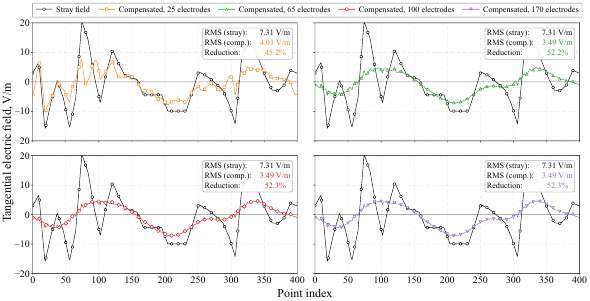}
\caption{\protect\justifying Tangential component of the electric field at different points for $N \in \{25,65,100,170\}$ compensation electrodes. The blue line is stray field before compensation at 45 points reported in Ref.~\cite{tabakov2015Assembling}. We use linear interpolation to obtain values in 400 points. Moreover, in areas where field was not measured, we use a constant value. Orange, green, red and purple lines represent the field after compensation for \{25,65,100,170\} electrodes respectively. Every 10th point is marked.}
\label{fig:comps_stray}
\end{figure*}

The compensated tangential stray field obtained for different numbers of electrode pairs $N$ is shown in Fig.~\ref{fig:comps_stray}.  The standard deviation for $N = 25$ equals 4.0 V/m. For other values of $N$, it is 3.4 V/m. It can be seen that $N = 65$ electrode pairs are sufficient to achieve a substantial reduction of the stray field.  To verify this result, we increased the number of electrode pairs up to $N = 170$. However, the compensated field remained nearly unchanged, indicating that adding more electrodes provides only marginal improvement. It is worth noting that, even for $N = 170$, the compensated stray field does not vanish completely at all points. This residual field arises from the voltage normalization constraint: in our optimization, the maximum electrode voltage is limited due to normalization to approximately $\pm22$~V. If this normalization is removed, the linear system admits an exact solution with complete compensation of the stray field. However, the required electrode voltages become excessively large, making such a solution impractical for experimental implementation. Finally, the perturbation of the trapping potential caused by the loading aperture can be mitigated by optimizing the geometry of the RF electrodes, as demonstrated in Ref.~\cite{moehring2011design}.

\subsection{Readout and manipulation with optical clocks}
\begin{figure}[t]
    \centering
    \subfloat[\label{fig:Ca_levels}]{
        \includegraphics[width=0.52\linewidth]{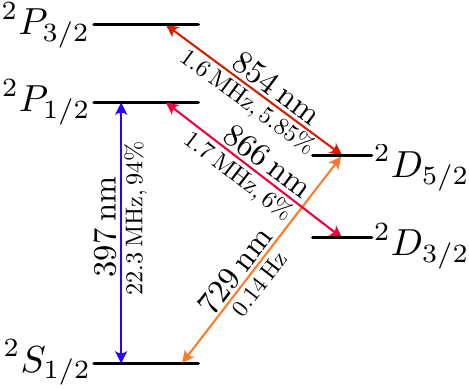}
    }
    \subfloat[\label{fig:Ca_beams}]{
        \includegraphics[width=0.45\linewidth]{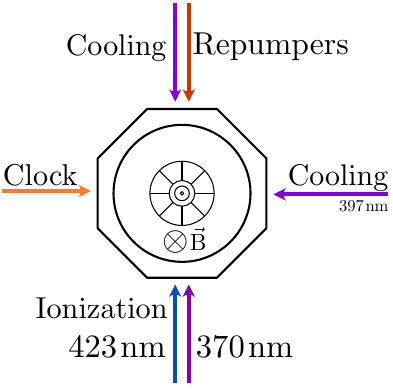}
    }
    \caption{\protect\justifying (a) Partial energy structure of the $\rm{Ca}^+$ ion showing transitions used for optical ion clock operation. (b) Possible experimental beam configuration. Two perpendicular cooling beams cover all motional projections during Doppler cooling procedure. One cooling beam and repumping beams propagate along the photoionization direction, simplifying the alignment.}
    \label{levels}
\end{figure}

We propose $\rm{Ca}^+$ ions as a simple, well-studied candidate for experimental implementation of ring-trap multi-ion clocks. Convenient energy structure, including near-UV 397 nm dipole transition with 22.3 MHz linewidth, allows for effective Doppler cooling (Fig.\ref{fig:Ca_levels}). Together with 866-nm and 854-nm repumping lasers, fast and high-fidelity state preparation can be done. The high-Q clock transition at 729 nm shows almost vanishing readout errors and one of the best instabilities \cite{dietze2026EntanglementEnhanced}. 
Even isotope $^{40}\rm{Ca}^+$ is good for trapping large ion crystals due to its high natural abundance of $\sim 97 \%$. The linear magnetic sensitivity of the clock transition could be removed by averaging over symmetric pairs of Zeeman transitions \cite{zhang2026LiquidNitrogenCooled}. On the other hand, for the $^{43}\rm{Ca}^+$ isotope, clock transition is naturally magnetically insensitive, but careful ion loading with isotope-selective photoionization \cite{lucas2004Isotopeselective} and special crystal purification techniques is required \cite{kuo2024Preparing}.

One of convenient laser beam and magnetic field configurations is shown in Fig.\ref{fig:Ca_beams}. To exploit the rotational symmetry of the ion ring, the quantization field is aligned along the symmetry axis of the crystal, perpendicular to the ion plane. This choice ensures that the electric-field-gradient tensor has the same projection onto the quantization axis for every ion, yielding identical electric quadrupole shifts in the ideal trap. Two orthogonal cooling beams are used to provide complementary projections onto the motional modes around the ion ring. Their relative frequency and polarization should be chosen to suppress coherent interference in the overlap region. The theoretical Doppler cooling limit for radial COM modes for the calculated radial frequencies is $\bar n \approx 3$ phonons per mode. One of the cooling beams together with repumping beams can be aligned using counter-propagating ionization beams.
The clock laser, directed perpendicular to the magnetic field with polarization along it, allows effective excitation $\Delta m_j\pm 1$ transitions for interrogation over a symmetrical Zeeman pair. Its initial alignment is also simplified due to overlap with the cooling beam.

For a conservative estimate of the required laser power, we use the following upper bound for the effective saturation intensity:
\begin{equation}
	I_{\rm{s}}\approx \hbar c\frac{\gamma}{2\pi}k^3,
\end{equation}
with wavevector $k$ and natural linewidth of considered transition $\gamma$. We consider Gaussian and flat-top beam shapes. The required optical power for intensity uniformly distributed over the beam cross-section is $P_{\rm{F}} = I_{\rm{s}} \times S$. For a Gaussian beam
\begin{equation}
	I(x,y) = I_0\cdot\exp\left(-\frac{2x^2}{\omega_x^2}-\frac{2y^2}{\omega_y^2}\right)
\end{equation} 
we have to ensure that intensity is sufficient for outermost ions:
\begin{equation}
	I_{\rm{out}} = I_{\rm{s}} = I_0\cdot \exp\left(-\frac{2r^2}{\omega_x^2}\right).
\end{equation}
The power could be calculated as
\begin{equation}
	P_{\rm{G}} = I_{\rm{s}}\frac{\pi\omega_x\omega_y}{2}\cdot\exp(\frac{2r^2}{\omega_x^2}).
\end{equation}
Here, $\omega_x$ and $\omega_y$ are the semi-major and semi-minor ellipse axes, respectively. An ion ring with radius of $r \sim 600 \hspace{2pt}\mu$m contains $\sim 400$ ions at $10 \hspace{2pt}\mu$m spacing corresponding to $\sim 0.9^\circ$ angular spacing. To ensure full coverage of the ion crystal with sufficient margin, a beam size of $1600\times 80$ $\mu$m (diameters) was chosen for both profile types. The Rayleigh length associated with the tightly focused axis is approximately 6 mm for repumping beams and larger for cooling beams, which is sufficient to maintain an essentially constant beam profile over the entire ion crystal. For these parameters less than $3\times 10^{-5}$ of Gaussian-beam power will hit the trap surface. The clock beam power requirement can be estimated relying on $\tau_\pi = 4$ ms pulse duration used in Hyper-Ramsey interrogation sequence in Ref.~\cite{zhang2026LiquidNitrogenCooled}. The saturation intensity can be rewritten as
\begin{equation}
    I_{sq} = \frac{\hbar c\pi k^3}{8\gamma\tau_\pi^2\sigma^2},
\end{equation}
where $\sigma = 1/2$ is a dimensionless angular coupling factor.
The resulting values are summarized in Table \ref{powers} and can readily be achieved, despite the requirement for a large beam size. 

\begin{table}[t]
\centering
\vspace{-3pt}
\renewcommand{\arraystretch}{1.2} 
\setlength{\tabcolsep}{2pt}

\begin{tabular}{lccc}
\hline\hline
 & \multicolumn{3}{c}{Wavelength} \\
\cline{2-4}
Parameter & 397 nm & 866, 854 nm & 729 nm \\
\hline
Saturation intensity, $\mu$W/mm$^{2}$
& 2790 
& 20 
& 1.2  \\

Gaussian beam power, $\mu$W 
& 430
& 3.1
& 0.12 \\

Flat-top beam power, $\mu$W 
& 280 
& 2.0 
& 0.18  \\

\hline\hline
\end{tabular}
\caption{\protect\justifying Conservative estimation of the optical powers required for multi-ion clock operation.}
\label{powers}
\end{table}

\begin{figure}[t]
    \centering
    \subfloat[\label{fig:objective1}]{
        \includegraphics[width=0.98\linewidth]{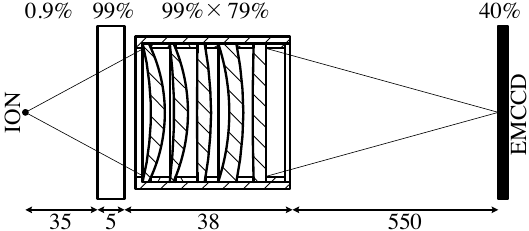}
    }
    \hfill
    \vspace{-0.2 cm}
    \subfloat[\label{fig:objective2}]{
        \includegraphics[width=1.0\linewidth]{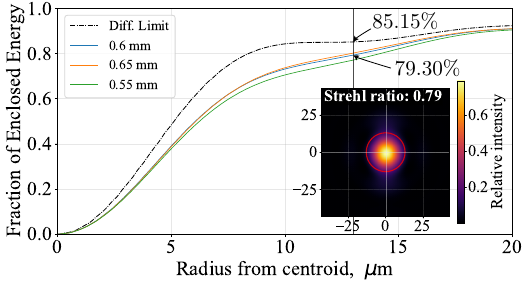}
    }
    \caption{\protect\justifying (a) Principal readout scheme (not to scale), dimensions are in mm. (b) Encircled-energy versus radius plot for the readout objective. For an Airy disk with $r = 13 \hspace{2pt}\mu$m we obtain $>75\%$ for ion rings of $r\in\{500,700\} \hspace{2pt}\mu$m. Different line colors correspond to different ion displacements relative to principal optical axis of the objective. The inset shows Huygens point-spread function for 600 $\mu$m off-axis light. Axes tick marks are in $\mu$m. The calculated Strehl ratio is 0.79, indicating $< \lambda/13$ RMS wavefront error. }
    \label{obj}
\end{figure}

The most challenging configuration for ion readout places the objective outside the vacuum chamber, where the long working distance limits the achievable numerical aperture. Nevertheless, this arrangement offers the simplest technical implementation. To suppress the BBR shift, a cryogenic environment is desirable \cite{zhang2017Observation, huang2022LiquidNitrogenCooled, zhang2026LiquidNitrogenCooled}. A typical viewport-ion spacing would be 30-40 mm \cite{pagano2019Cryogenic}. The readout objective must simultaneously satisfy several demanding requirements: efficient fluorescence collection during the standard 1 ms readout interval, a sufficiently large field of view for simultaneous imaging of the entire ion ring, and high image quality across the full field through correction of both intrinsic optical aberrations and aberrations introduced by the vacuum viewport. The largest commercially available EMCCD sensor has a resolution of $1024\times1024$ with a pixel size of $13 \hspace{2pt}\mu$m. With a 9$\times$ magnification, the ion ring would produce an approximately 11 mm diameter image on the camera, providing near-maximal inter-ion spacing on the sensor for reduced crosstalk while retaining a modest margin within its area. We estimate the minimal clear aperture by the condition that an ion should be projected to a single pixel in the image plane. Therefore, with an Airy radius of $r = 13/(2\cdot9)$ $\mu$m in the object plane we find
\begin{equation}
	D = 1.22\frac{\lambda f}{r} \approx 32 \hspace{2pt}\rm{mm},
\end{equation}
where $\lambda = 397$ nm and $f = 48$ mm (taking into account objective-viewport spacing). We then choose 35 mm lens diameter and present custom-designed five-element N-BK7 objective with working distance $\approx 45$ mm that is corrected for a 5-mm fused-silica window (Fig. \ref{obj}). We intentionally decreased NA to 0.19 to improve the image quality for various ring diameters and achieve $>75\%$ Airy-disk encircled energy for crystals with $r\in\{500,700\} \hspace{2pt}\mu$m. Each ion would appear as an approximately $2\times2$-pixel square, with a separation of about 6 pixels between neighboring ions. Lens shapes are given in Table \ref{lens}.

Assuming 40\% quantum efficiency for EMCCD sensor, 99\% viewport and objective transmission with 75\% encircled power, the detection efficiency would be 0.28\% of the scattered light, which gives 31 photons/ms for the bright state. For a dark state count rate at 1 photon/ms (usually non-zero due to diffused cooling laser light), a bright state count rate of 27 photons/ms leads to $\sim 10^{-5}$ readout errors, leaving space for objective manufacturing and positioning imperfections as well as unaccounted losses in the readout optical path.

\begin{table}[h]
\centering
\begin{tabular}{c|r|r|c|r|}
\rowcolor[HTML]{DDDDDD}[2pt][2pt] 
\hline
\multicolumn{1}{|l|}{} & Radius & Thickness & Glass & Semi-Diameter \\ \hline
\multicolumn{1}{|l|}{1}  & $\infty$   & 35.0      &       & ION           \\ \hline
\rowcolor[HTML]{CBCEFB}[2pt][2pt] 
\multicolumn{1}{|l|}{2}  & $\infty$   & 5.0       & F\_SILICA & 17.5          \\ \hline
\multicolumn{1}{|l|}{3}  & $\infty$   & 6.0     &       & 17.5          \\ \hline
\rowcolor[HTML]{CBCEFB}[2pt][2pt]  
\multicolumn{1}{|l|}{4}  & -34.6    & 3.0       & N-BK7     & 17.5          \\ \hline
\multicolumn{1}{|l|}{5}  & $-32.6$    & 2.0       &       & 17.5          \\ \hline
\rowcolor[HTML]{CBCEFB}[2pt][2pt]  
\multicolumn{1}{|l|}{6}  & $-50.3$   & 3.0       & N-BK7     & 17.5          \\ \hline
\multicolumn{1}{|l|}{7}  & $-38.7$    & 2.0       &       & 17.5          \\ \hline
\rowcolor[HTML]{CBCEFB}[2pt][2pt]  
\multicolumn{1}{|l|}{8}  & $-80.1$   & 3.0       & N-BK7     & 17.5          \\ \hline
\multicolumn{1}{|l|}{9}  & $-49.6$    & 4.0       &       & 17.5          \\ \hline
\rowcolor[HTML]{CBCEFB}[2pt][2pt]  
\multicolumn{1}{|l|}{10} & $-183.1$    & 3.0       & N-BK7     & 17.5          \\ \hline
\multicolumn{1}{|l|}{11} & $-70.4$    & 2.0       &       & 17.5          \\ \hline
\rowcolor[HTML]{CBCEFB}[2pt][2pt]  
\multicolumn{1}{|l|}{12} & $663.7$    & 3.0       & N-BK7     & 17.5          \\ \hline
\multicolumn{1}{|l|}{13} & $-125.3$  & 550.0     &       & EMCCD         \\ \hline
\end{tabular}
\caption{\protect\justifying Lens data of the ion fluorescence readout objective. The objective is designed for a wavelength of 397 nm with a working distance of 46 mm. The maximum object-space numerical aperture is 0.35 and can be reduced to 0.19 by adjusting the entrance aperture to accommodate ion rings of different diameters. Values are given in mm. }
\label{lens}
\end{table}

\section{Systematic shifts in multi-ion optical clock} \label{sec:error_budget}
In this section, we discuss the main frequency shifts relevant to the proposed multi-ion clock and their contributions to its uncertainty budget. We do not consider blackbody radiation (BBR) shifts or collisions with background gas, as these effects are not specific to multi-ion operation and have been extensively studied in single-ion clocks~\cite{zhang2017Evaluation, hankin2019Systematic, huang2022LiquidNitrogenCooled, lindvall202588}.
\subsection{Quadrupole shift}
The quadrupole shift arises from electric-field gradients produced by the trapping fields, neighboring ions (space charge), and stray fields. In an ideal, rotationally symmetric harmonic trapping potential, the DC trapping field produces the same quadrupole shift for all ions equally spaced around a ring. Different pertirbations can introduce spatially varying field gradients~\cite{lin2009Largescale, home2011normal}, which will lead to increased uncertainty. The first-order contribution from the RF field averages to zero over an RF cycle.

Space-charge-induced quadrupole shifts are a major challenge for scalable multi-ion clocks. In a linear chain, ions at different positions experience different electric-field gradients, leading to inhomogeneous frequency shifts and line broadening~\cite{arnold2015Prospects, aharon2019Robust}. In an ideal ring crystal, rotational symmetry of the ion crystal makes the space-charge-induced quadrupole shift identical for all ions. With an appropriate orientation of the quantization axis, the resulting quadrupole shift is therefore common to all ions and does not cause inhomogeneous broadening.

This advantage relies on preserving the circular symmetry and uniform ion spacing through stray-field compensation. If the stray fields remain stable over the measurement period, compensation can be performed before clock operation and maintained throughout data acquisition.

Following Refs.~\cite{itano2000Externalfield, arnold2015Prospects, aharon2019Robust}, the quadrupole shift of the $i$th ion can be written as

\begin{equation}
\label{eq:space_charge}
\Delta \nu_Q^{i} =
\frac{\Theta}{h}
\frac{ J(J+1) - 3m_J^2 }
{J(2J-1)}
f_{i}(\alpha, \beta, \gamma),
\end{equation}
where $\Theta$ is the electric quadrupole moment of the clock state, $J$ and $m_J$ are its electronic angular momentum and magnetic quantum number. The factor $f_i(\alpha,\beta,\gamma)$ describes the local electric-field gradient and its orientation relative to the quantization axis, as specified by the angles $\alpha$, $\beta$, and $\gamma$~\cite{arnold2015Prospects, aharon2019Robust}. For the clock state of $^{40}\mathrm{Ca}^{+}$, the measured quadrupole moment is $\Theta = 1.83(1) \ e a_0^2$~\cite{roos2006Designer}.

\begin{figure}[t]
    \centering
    \subfloat[\label{fig:quadrupole_shift_ring}]{
        \includegraphics[width=1.0\linewidth]{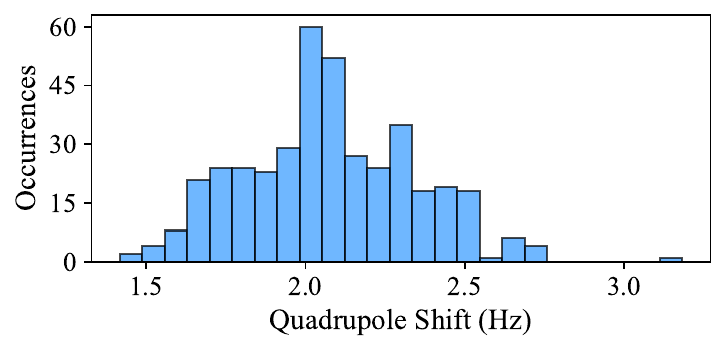}
    }
    \hfill
    \vspace{-0.5 cm}
    \subfloat[\label{fig:quadrupole_shift_linear}]{
        \includegraphics[width=1.0\linewidth]{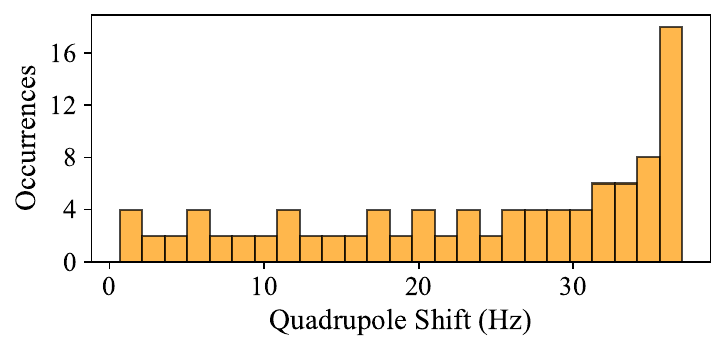}
    }
    \caption{\protect\justifying (a) Quadrupole shift for the~$m_J = 1/2$ state for 400 ions in a ring surface trap. The distribution has a mean shift~$2.07$~Hz and a standard deviation of 0.27~Hz. (b) Simulation of the quadrupole shift in a linear trap for 100 ions. The distribution has a mean shift~$24$ Hz and uncertainty~$11$~Hz.  }
    \label{fig:quadupole_shift}
\end{figure}

Using the approach of Ref.~\cite{arnold2015Prospects}, we calculate the space-charge-induced quadrupole shifts in a ring crystal containing 400 ions. For an ideal ring with uniform ion spacing, the shifts are identical under the symmetry conditions described above. Stray fields perturb the equilibrium positions, break this equivalence, and introduce a spread of quadrupole shifts. To quantify this effect and assess the compensation method presented in Sec.~\ref{sec:compensation}, we numerically determine the equilibrium positions of 400 ions for the stray-field configurations before and after compensation shown in Fig.~\ref{fig:comps_stray}.

We consider a configuration of the trap with $N = 65$ electrode pairs. The Fig.~\ref{fig:comps_stray} shows, that it should be enough. For each ion we evaluate the total electric field at its location. The field has three contributions: (a) the Coulomb field of all other ions; (b) the stray field; and (c) the field produced by the trap electrodes. The Coulomb term is computed numerically from the instantaneous ion distribution; the tangential profile of the stray field is taken from Ref.~\cite{tabakov2015Assembling} and linearly interpolated; the trap field is computed with the Python package \texttt{electrode}. The radial and axial confinement is much stiffer than the tangential one, so the ions are effectively constrained to the ring and only the tangential field component is relevant for the equilibrium configuration. The objective function therefore takes 400 parameters -- the angular position of each ion -- and returns the 400
tangential field components, one per ion. The resulting square system is solved with \texttt{scipy.optimize.least\_squares}. At the equilibrium configuration the tangential field vanishes at every ion.

We then use these equilibrium positions to calculate the quadrupole shift of each ion. Figure~\ref{fig:quadrupole_shift_ring} shows the distribution of space-charge-induced shifts before stray-field compensation (zero voltages on DC electrodes). For compensation we use voltages calclulated in Sec.~\ref{sec:compensation} for choosen configuration. After stray field compensation, the simulated ion spacing is nearly uniform, and the estimated fractional uncertainty associated with the residual quadrupole-shift inhomogeneity is below $10^{-19}$.

For comparison, we perform the same calculation for a linear chain of 100 ions in an ideal harmonic trap with an axial secular frequency of $\nu_a = 80$~kHz~\cite{leibfried2003Quantum}. We first determine the equilibrium positions and then calculate the quadrupole shift of each ion. The resulting distribution is shown in Fig.~\ref{fig:quadrupole_shift_linear}. Even before stray-field compensation, the spread of quadrupole shifts in the ring is approximately two orders of magnitude smaller than in the ideal linear chain. This comparison does not include trap anharmonicities, which can further increase the shift inhomogeneity.

Residual quadrupole shifts can be further suppressed by continuous dynamical decoupling during interrogation. Two approaches have been demonstrated. The resonant scheme used in Refs.~\cite{shaniv2019Quadrupole, akerman2025Operating, sokolik2026Direct} offers a relatively simple implementation but requires magnetic-field calibration between interrogation cycles. A more general, two-stage scheme was proposed in Ref.~\cite{aharon2019Robust} and demonstrated experimentally in Ref.~\cite{pelzer2024Multiion}. Its additional control parameters enable simultaneous suppression of the linear Zeeman and quadrupole shifts.

Reference~\cite{pelzer2024Multiion} estimates a quadrupole-shift suppression factor of $3\times10^{-2}$. We use this factor to estimate the residual contributions to the uncertainty budget in Table~\ref{tab:budget} for a linear trap and ring trap before compensation of stray fields. The quadrupole contribution remains a limiting factor for the linear-chain clock, whereas the ring geometry, combined with stray-field compensation, is compatible with a fractional clock uncertainty at the $10^{-18}$ level.

The simultaneous suppression of the linear Zeeman shift is particularly valuable for $^{40}\mathrm{Ca}^{+}$, which has no hyperfine structure and is therefore highly sensitive to magnetic-field fluctuations on the clock transitions considered here.
\subsection{Excess Micromotion}
Excess micromotion occurs when ions are displaced from the RF null. In conventional linear Paul traps, axial micromotion arises from two distinct mechanisms: residual AC fields that leak onto the DC endcap electrodes~\cite{berkeland1998Minimization}, and from fringe fields near the ends of finite RF electrodes. In the ideal ring architecture, the RF null forms a closed loop that follows the ion crystal, and rotational symmetry eliminates the tangential RF field.

Realizing this advantage requires a high degree of rotational symmetry. Radial stray fields can displace ions from the RF null and produce excess micromotion. Their effects can be reduced through compensation electrodes or operation at a magic RF drive frequency~\cite{berkeland1998Minimization, arnold2015Prospects}. At this frequency, the micromotion-induced second-order Doppler and AC Stark shifts cancel. Such cancellation requires a negative differential static scalar polarizability, $\Delta\alpha < 0$.

For $^{40}\mathrm{Ca}^{+}$, the magic RF angular frequency has been measured as $\Omega_{\mathrm{magic}} = 2\pi \times 24.801(2)$~MHz~\cite{safronova2011Blackbody, huang2019Ca}. Other suitable candidates include $^{43}\mathrm{Ca}^{+}$~\cite{arora2007Blackbodyradiation}, $^{88}\mathrm{Sr}^{+}$~\cite{jiang2008Electric, dube2014HighAccuracy}, $^{138}\mathrm{Ba}^{+}$~\cite{sahoo2009Comparative}, $^{176}\mathrm{Lu}^{+}$~\cite{kozlov2014Optical, arnold2015Prospects}, and $^{225}\mathrm{Ra}^{+}$~\cite{sahoo2009Comparative}.

\begin{table}[t]
\centering
\vspace{-3pt}
\renewcommand{\arraystretch}{1.2} 
\setlength{\tabcolsep}{2pt}

\begin{tabular}{lccc}
\hline\hline
 \multirow{3}{*}{Effect} & \multicolumn{3}{c}{Uncertainty ($10^{-18}$)} \\
\cline{2-4}
 &  \multirow{2}{*}{Linear trap}   & \multicolumn{2}{c}{Ring trap}   \\
 & & \makecell{before comp.}  & after comp. \\
\hline
Quadrupole shift
&  834 
&  19.5 
& $< 0.1$
 \\
Excess micromotion
&  \multicolumn{3}{c}{0.5, \cite{huang2019Ca}}
 \\

Blackbody radiation 
&   \multicolumn{3}{c}{0.4, \cite{zhang2026LiquidNitrogenCooled}}  
  \\
Quadratic Zeeman
&  \multicolumn{3}{c}{0.1, \cite{zhang2026LiquidNitrogenCooled}} 
  \\
\hline
Total
 &  834
& 19.5 & $< 0.7$
  \\
\hline\hline
\end{tabular}
\caption{\protect\justifying Uncertainties of the fractional frequency shifts for the linear Paul trap and the ring trap. For the most significant homogeneous shifts (such as the blackbody radiation and quadratic Zeeman shifts), we adopt the uncertainties reported in the literature for the single-ion regime. }
\label{tab:budget}
\end{table}

\section{Discussion}
We have proposed a ring surface-trap architecture for optical clocks containing hundreds of ions. The rotational symmetry of the trapping configuration reduces the spatial variation of electric-field gradients and suppresses tangential micromotion in the ideal geometry. These properties address two major challenges in scaling trapped-ion clocks: inhomogeneous quadrupole shifts and micromotion-related frequency shifts.

To assess the effect of symmetry-breaking perturbations, we numerically determine the equilibrium positions of a 400-ion ring crystal in the presence of external electric fields, both before and after stray-field compensation. This allows us to evaluate the space-charge-induced quadrupole shifts using the perturbed crystal configurations. For the field configurations considered here, the spread of quadrupole shifts is approximately two orders of magnitude smaller than in the reference 100-ion linear chain even before compensation. After compensation, the ion spacing becomes nearly uniform, substantially reducing the residual shift inhomogeneity.

Stray-field control is therefore a key experimental requirement. Tangential fields are particularly important because they directly perturb the ion spacing and break the rotational symmetry of the crystal. Using 65 electrode pairs, the proposed compensation method reduces the tangential stray field by approximately 50\% for the configuration studied here. The simulations of equilibrium positions of ions connect this stray field reduction to the resulting improvement in crystal uniformity and quadrupole-shift homogeneity. The compensation model can also be extended to all three spatial directions without a substantial increase in the number of electrodes.

The ring geometry provides a favorable starting point for further suppression of systematic shifts. Continuous dynamical decoupling can reduce residual quadrupole and Zeeman shifts~\cite{pelzer2024Multiion, sokolik2026Direct}, magic RF operation can suppress micromotion-induced frequency shifts~\cite{huang2022LiquidNitrogenCooled}, and cryogenic operation can reduce the BBR shift~\cite{zhang2026LiquidNitrogenCooled}. Combined with these techniques, the proposed architecture offers a route toward a fractional frequency uncertainty at the $10^{-18}$ level. For 400 independently interrogated ions, the ideal quantum-projection-noise scaling gives a 20-fold improvement in frequency stability relative to a single ion, assuming the same interrogation time and contrast.

The architecture could be extended to thousands of ions by increasing the trap size and the number of compensation electrodes. Such scaling would require suitable electrode-control strategies, as explored for large surface traps in Ref.~\cite{malinowski2023How}, together with interrogation and readout adapted to the larger crystal, potentially including spatially resolved methods~\cite{tabakov2015Assembling}. Although substantial engineering challenges remain, the ring geometry provides a promising route to optical clocks with large trapped-ion ensembles.

\begin{acknowledgments}
We thank Kelin Gao, Hua Guan, and Zhiqiang Zhang for helpful discussion. This work is supported by National Natural Science Foundation of China grant no. 92265208 and Beijing Foreign High-level Talent Program grant no. C2024003.
\end{acknowledgments}

 \bibliography{ring_v3}
\appendix

\end{document}